\documentclass{kluwer}

\newcommand{\ang}{\AA\ }
\newcommand{\gapprox}{\lower.4ex\hbox{$\;\buildrel >\over{\scriptstyle\sim}\;$}}
\newcommand{\lapprox}{\lower.4ex\hbox{$\;\buildrel <\over{\scriptstyle\sim}\;$}}

\usepackage{epsfig}

\begin{document}
\begin{article}
\begin{opening}

\title{		Review of Coronal Oscillations - An Observer's View}
\runningtitle{	CORONAL OSCILLATIONS}

\author{	Markus J. \surname{Aschwanden}\email{aschwanden@lmsal.com}}
\institute{	Lockheed Martin Advanced Technology Center,
                Solar \& Astrophysics Laboratory,
                Dept. L9-41, Bldg.252,
                3251 Hanover St.,
                Palo Alto, CA 94304, USA}

\runningauthor{\surname{M. ASCHWANDEN}} 

\begin{abstract}
Recent observations show a variety of oscillation modes in the corona. 
Early non-imaging observations in radio wavelengths showed 
a number of fast-period oscillations in the order of seconds, 
which have been interpreted as fast sausage mode oscillations. 
TRACE observations from 1998 have for the first time revealed 
the lateral displacements of fast kink mode oscillations, with 
periods of $\approx$3-5 minutes, apparently triggered by nearby flares 
and destabilizing filaments. Recently, SUMER discovered with 
Doppler shift measurements loop oscillations with longer periods 
(10-30 minutes) and relatively short damping times in hot (7 MK) loops, 
which seem to correspond to longitudinal slow magnetoacoustic 
waves. In addition, propagating longitudinal waves have also 
been detected with EIT and TRACE in the lowest density scale 
height of loops near sunspots. All these new observations seem to 
confirm the theoretically predicted oscillation modes and 
can now be used as a powerful tool for ``coronal seismology'' 
diagnostic. 
\end{abstract}

\end{opening}

\section{Introduction}

Some 30 years ago, solar radio astronomers discovered a host
of solar radio events with periodic or quasi-periodic pulses,
with typical periods in the range of $P\approx 0.5-5$ s
(e.g. Rosenberg 1970). The
evidence for coronal oscillations was entirely based on the
periodicity observed in time profiles of the fluctuating
radio flux, which occurred at decimetric frequencies 
and thus were known to originate in the corona.
Imaging observations of such pulsating radio events
were largely not available, or only in form of 1-dimensional
brightness maps with poor spatial resolution. Nevertheless,
these radio observations stimulated some theoretical work on
MHD oscillations (e.g. Roberts et al. 1984) and the 
observed period range of $P\approx 0.5-5$ s was correctly 
identified in terms of {\sl fast sausage mode} oscillations. However, 
little progress has been made over the next 20 years, mainly
because of the lack of high-resolution imaging observations. 

Some 3 years ago, when high-resolution imaging observations on
sub-arcsec\-ond scale became available with the {\sl Transition
Region and Coronal Explorer (TRACE)}, the transverse motion
of {\sl fast kink MHD mode} oscillations in coronal loops was 
for the first time spatially 
resolved (Aschwanden et al. 1999). Additionally, the {\sl Solar
Ultraviolet Measurements of Emitted Radiation (SUMER)} instrument
onboard the {\sl Solar and Heliospheric Observatory (SoHO)}
discovered recently Doppler shifts of {\sl slow mode MHD oscillations} 
in coronal loops (Wang et al. 2002a,b). 
All these instrumental capabilities permit us now
to measure the periods, damping times, spatial displacements, 
velocities, and loop geometries in great detail, which 
rekindled theoretical work on MHD oscillations and
waves to a great deal. It opened up the new field of {\sl ``coronal
seismology''} (Roberts et al. 1984), 
which is expected to reveal fundamental physical
properties of the solar corona with similar sensitivity as the
discipline of {\sl ''helioseismology''} provides for the solar 
interior. 

\section{Coronal Oscillations in Radio}

The physical mechanisms of radio pulsation events can be
subdivided into three categories: (1) MHD eigen-mode
oscillations, (2) limit cycles of nonlinear dissipative
systems, and (3) modulation by a quasi-periodic particle
acceleration or injection mechanism (for a review, see
Aschwanden 1987a). In Table I-III we compile radio observations 
of coronal oscillations in three different period ranges 
(Table I: fast periods $P<0.5$ s; 
 Table II: short periods $0.5<P<5$ s; 
 Table III: long periods $P>5$ s).
The subdivision into these three period ranges is a little
bit arbitrary, but is believed roughly to separate 
three (or more) different physical processes. 
We list in Tables I-III the observed periods $P$, 
the radio frequencies $\nu$, 
and the spatial scale (either the measured spatial
source size or the instrumental resolution).

\subsection{Very Fast Radio Oscillations}

\begin{table}
\begin{tabular}{llll} \hline
Observer 		& Frequency     & Period        & Spatial \\
	 		& $\nu$ [MHz]   & P [s]         & Scale \\
\hline
Droege (1967) 		& 240, 460      & 0.2-1.2       &\\
Elgaroy and Sveen (1973) & 500-550	& 0.1-0.3	&\\
Tapping (1978)		& 140		& 0.06-5	&\\
Pick and Trottet (1978) & 169		& 0.37		&5'-7' (Nan\c{c}ay)\\
Gaizauskas and Tapping (1980) & 10,700	& 0.4		&2.7' (Algonquin)\\
Takakura et al. (1983)	& 22,000, 44,000 & 0.25-0.33	&\\
Costa and Kaufmann (1986) & 90,000	& 0.15		&\\
Kaufmann et al. (1986)	& 90,000	& 0.06		&\\
Elgaroy (1986)		& 305-540	& 0.05-0.15	&\\
Li et al. (1987)	& 327		& 0.3		&\\
Fu et al. (1990)	& 5380-6250	& 0.16-0.18	&\\
Kurths et al. (1991)	& 234-914	& 0.07-5.0	&\\
Fleishman et al. (1994)	& 2500, 2850	& 0.07-0.08	&\\
Chernov et al. (1998)	& 164-407	& 0.2		&5'-7' (Nan\c{c}ay)\\
Makhmutov et al. (1998)	& 48,000	& 0.2-0.5	&1.9' (Itapetinga)\\
\hline
\end{tabular}
\caption{Coronal oscillations observed in radio (very fast periods: $P<0.5$ s)}
\end{table}

In Table I we compile radio observations with very fast (sub-second) periods, 
measured down to 50 ms. Such short periods are most likely not produced by a
MHD mode of an oscillating coronal loop, but rather by the nonlinear dynamics
of small-scale phenomena. If we assume a typical Alfv\'en speed of
$v_A \approx 1000$ km s$^{-1}$ in the solar corona, time periods of
$P\approx 0.05-0.5$ s would require spatial scales of $L\approx P \times v_A 
\approx 50-500$ km, which are way below diameters of typical active
region loops observed in EUV and soft X-rays. However, such small spatial
scales have been estimated from the spectral bandwidth of decimetric
millisecond spikes, which are believed to be a manifestation of fragmented 
energy release (Benz 1985). Moreover, elementary time structures of hard
X-ray emission have similar sub-second time scales (Aschwanden et al. 1995a),
which are believed to reflect the spatially and temporally
intermittent acceleration and injection of electrons from dynamic current
sheets operating in a {\sl bursty magnetic reconnection mode} (LeBoef et 
al. 1982; Priest 1985; Tajima et al. 1987; Kliem 1988, 1995). Numerical
MHD simulations have shown that non-steady reconnection operates in 
quasi-periodic cycles of tearing mode filamentation and coalescence
(Tajima et al. 1987; Karpen et al. 1995; Kliem et al. 2000). Recent physical
models of sub-second pulsations inferred from radio observations therefore
deal with quasi-periodic particle acceleration and injection
(Takakura et al. 1983; Costa \& Kaufmann 1986; Kaufmann et al. 1986;
Fleishman et al. 1994; Kliem et al. 2000).

Alternative interpretations, mostly published earlier, deal with 
a pulsating re\-gi\-me of ion beam instabilities (Zaitsev 1971), 
a pulsating regime of loss-cone instabilites (Chiu 1970; 
Zaitsev \& Stepanov 1975; Aschwanden \& Benz 1988), 
modulation of gyro-synchrotron emission by propagating 
whistler wave packets (Tapping 1978; Li et al. 1987; Chernov 1989), 
modulation by torsional Alfv\'en waves (Tapping 1983) caused by repetitive 
collapses of double layers (Tapping 1987), 
bounce times of trapped electrons (Elgaroy 1986),
or modulation of gyrosynchrotron emission by radial (fast-sausage)
MHD mode oscillations (Gaizaus\-kas \& Tapping 1980).
What is common to most of these dynamic scenarios is that the
pulsating regime corresponds to the limit cycle in phase space
(e.g. Weiland \& Wilhelmsson 1973)
of a nonlinear dissipative system, and thus may show minor (for linear
disturbances) or larger (for highly nonlinear conditions) deviations
from strict periodicity. The nonlinear properties of such systems
have been explored by measurements of the strange attractor dimension
(Kurths \& Herzel 1986, 1987; Kurths et al. 1991) and the bifurcation 
of periods on the (Ruelle-Takens-Newhouse) route to chaos
(Kurths \& Karlicky 1989).
Strange attractor dimension analysis of radio pulsation events
yielded $D\approx 2.5-3.5$, which means that a coupled equation
system of about 3-4 independent physical parameters is needed to
describe the nonlinear dissipative system (Kurths et al. 1991). 

Very fast oscillations with periods of $P<0.1$ s could also result
from heating and acceleration of minor ions by dissipation of high-frequency
(10-$10^4$ Hz) ion-cyclotron waves, as they have been employed to explain
the large perpendicular ion temperature anisotropies ($\approx 10^8$ K) 
inferred from the excessive line broadening of O VI lines by SoHO/UVCS
(Kohl et al. 1997, 1999). These signatures, however, have been observed 
in coronal holes, while fast radio pulsations are generally associated 
with flares, occurring at low latitudes. 

\begin{figure}[h]
\centerline{\epsfig{file=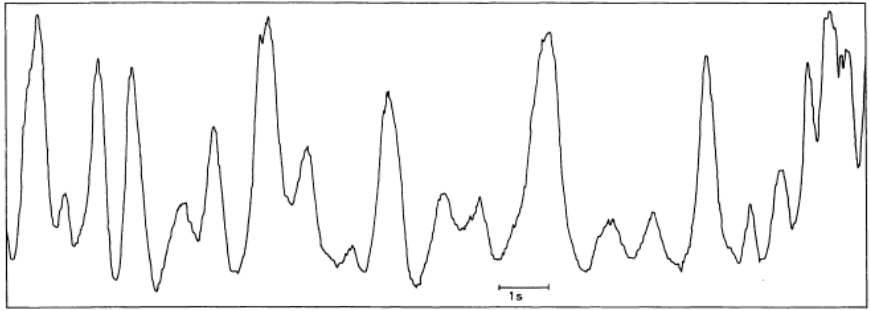,width=\textwidth}} 	
\caption{Microphotometric tracing of radio pulsations observed
on February 25, 1969, 09.55 UT, in the frequency range of
$\nu=160-320$ MHz at the Utrecht radio observatory. Note
that besides the fundamental period of $P_1\approx 3$ s, there
are also subharmonic pulse structures with periods of
$P_2\approx 1$ s visible (Rosenberg 1970).} 
\end{figure} 

\begin{figure}[h]
\centerline{\epsfig{file=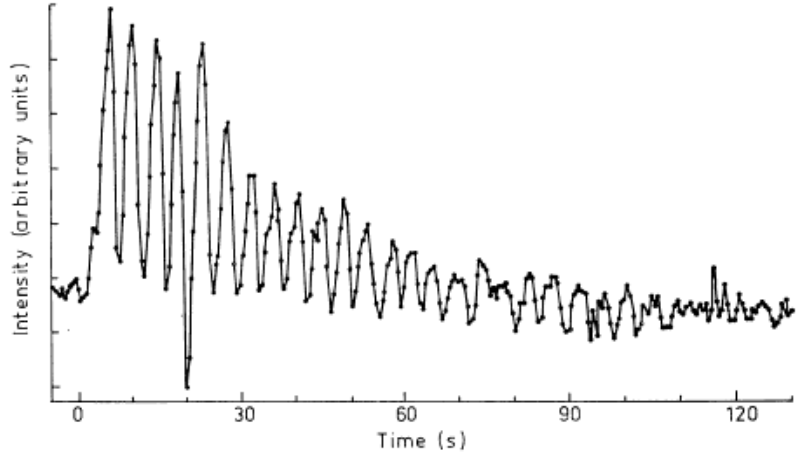,width=\textwidth}} 	
\caption{Microphotometric tracing of radio pulsations observed
on 1972 May 16, 03:14 UT, between 200 and 300 MHz, with the
Culgoora radio spectograph. The deep minimum at 20 s is produced
by the frequency calibration and time marker at 03:15 UT 
(McLean \& Sheridan 1973).}
\end{figure} 

\subsection{Short-Period Radio Oscillations}

\begin{table}
\begin{tabular}{llll} \hline
Observer 		& Frequency     & Period        & Spatial \\
	 		& $\nu$ [MHz]   & P [s]         & Scale \\
\hline
Droege (1967) 		& 240, 460      & 0.2-1.2       &\\
Abrami (1970, 1972)	& 239		& 1.7-3.1	&\\
Gotwols (1972)		& 565-1000	& 0.5		&\\
Rosenberg (1970)	& 220-320	& 1.0-3.0	&\\
De Groot (1970)		& 250-320	& 2.2-3.5	&\\
McLean et al. (1971)	& 100-200	& 2.5-2.7	&\\
Rosenberg (1972)	& 220-320	& 0.7-0.8	&\\
McLean and Sheridan (1973) & 200-300	& 4.20$\pm$0.01	&\\
Kai and Takayanagi (1973) & 160		& $<$1.0	&17' (Nobeyama)\\
Achong (1974)		& 18-28		& 4.0		&\\
Abrami and Koren (1978)	& 237		& -		&\\
Tapping (1978)		& 140		& 0.06-5	&\\
Pick and Trottet (1978) & 169		& 0.37, 1.7	&5'-7' (Nan\c{c}ay)\\
Elgaroy (1980)		& 310-340	& 1.1		&\\
Bernold (1980)		& 100-1000	& 0.5-5		&\\
Trottet et al. (1981)	& 140-259	& 1.7$\pm$0.5	&5' (Nan\c{c}ay)\\
Slottje (1981)		& 160-320	& 0.2-5.5	&\\
Sastry et al. (1981)	& 25-35		& 2-5		&\\
Kattenberg and Kuperus (1983) & 5000	& 1.5		&0.15' (Westerbork)\\
Wiehl et al. (1985)	& 300-1000	& 1-2		&\\
Aschwanden (1986, 1987b)& 300-1100	& 0.4-1.4	&\\
Aschwanden and Benz (1986) & 237, 650	& 0.5-1.5	&\\
Correia and Kaufmann (1987) & 30,000, 90,000 & 1-3	&\\
Kurths and Karlicky (1989) & 234	& 1.3, 1.5	&\\
Chernov and Kurths (1990) & 224-245	& 0.35-1.3	&\\ 
Zhao et al. (1990)	& 2840		& 1.5		&\\
Kurths et al. (1991)	& 234-914	& 0.07-5.0	&\\
Aschwanden et al. (1994)& 300-650	& 1.15, 1.8	&\\
Qin and Huang (1994)	& 9375		& 1.0-3.0	&\\
Qin et al. (1996)	& 2840, 9375, 15,000 & 1.5	&\\
Makhmutov et al. (1998)	& 48,000	& 2.5-4.5 	&1.9' (Itapetinga)\\
Kliem et al. (2000)	& 600-2000	& 0.5-3.0	&\\
\hline
\end{tabular}
\caption{Coronal oscillations observed in radio (short periods: $0.5 \le P \le 5$ s)}
\end{table}

\begin{figure}
\centerline{\epsfig{file=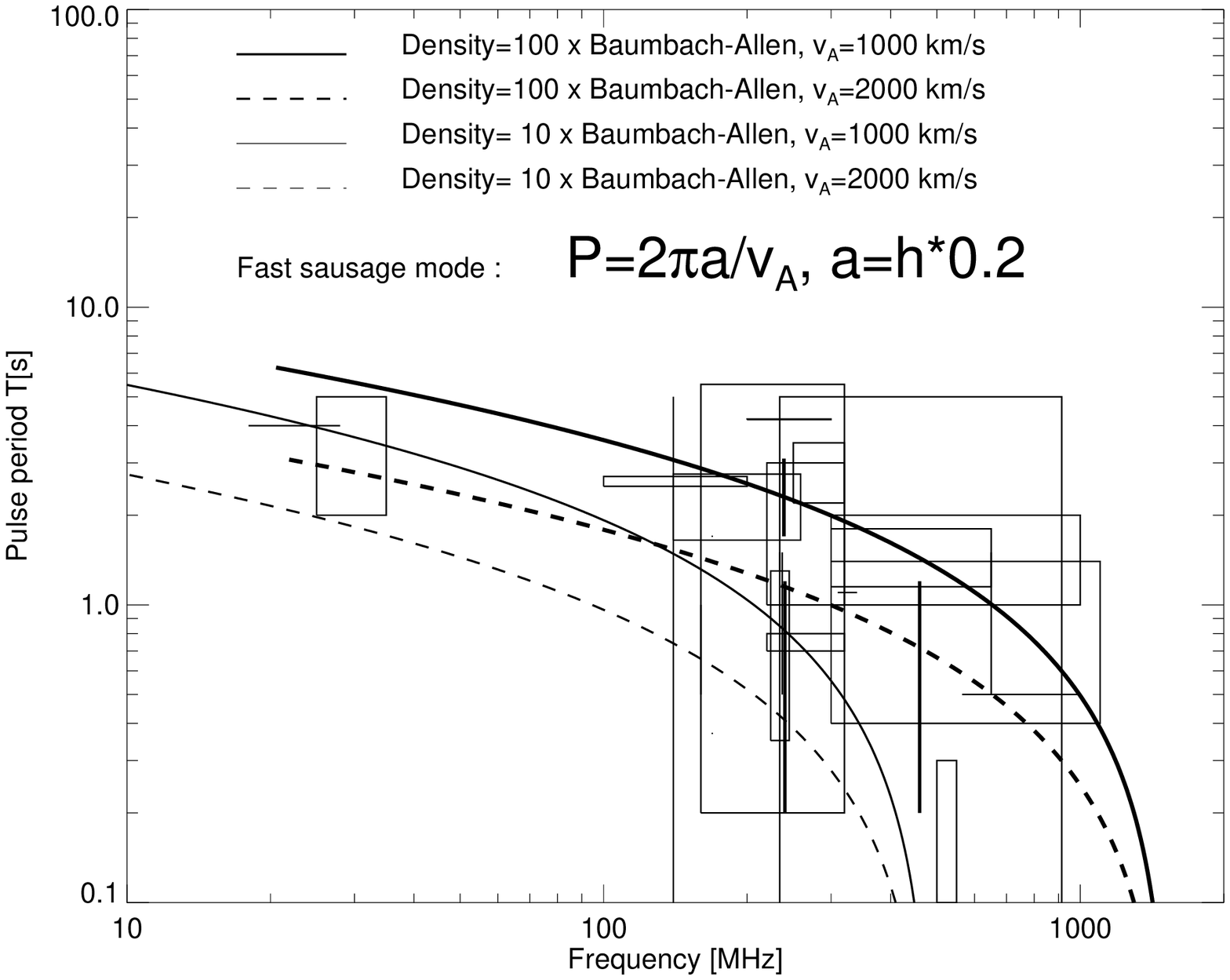,width=\textwidth}} 	
\caption{Measurements of radio pulsation periods $P$ as function of frequency $\nu$
(rectangle and bars) and expected theoretical periods of the fast sausage mode
as function of the plasma frequency (see model in text). Note the tendency
that the periods become longer at low frequencies.}
\end{figure} 

Radio pulsations with periods of order seconds are most common.
In Table II we compile a list of some 32 papers that contain observations
of radio pulsations with periods of $0.5<P<5.0$ s, detected at
metric, decimetric, and centimetric frequencies. Examples of such
observations are shown in Figs.1 and 2. Fig.1 shows the first detection
of radio pulsations that have been interpreted in terms of (fast-sausage)
MHD modes (Rosenberg 1970). The time profile shown in Fig.1 shows not only
the fundamental period of $P_1=3.0$ s, but also subpulses at the harmonic 
period, which has a period ratio of $P_1/P_2=5.3/1.8=3.0$ based
on the first-order Bessel function that characterizes the eigen-modes of
radial oscillations in a cylindrical fluxtube. Fig.2 shows a very periodic
pulsation with a period of $P=4.20\pm0.01$ s and a relatively long damping 
time that corresponds roughly to 10 pulse periods (McLean \& Sheridan
1973).

Radio pulsations with periods of order $0.5<P<5.0$ s have generally been
interpreted in terms of the fast sausage MHD mode, which is a standing
wave of a cylindrical fluxtube with radial cross-section variations.
The period of the fast sausage mode corresponds to the Alfv\'enic wave 
propagation across the loop cross-section $a$ (Roberts et al. 1984),
\begin{equation}
	P_{fast-sausage} 
	= 4 {\pi}^{3/2} a \left( { {\rho}_0 + {\rho}_e \over
		B_0^2 + B_e^2 } \right)^{1/2}
	\approx {2 \pi a \over v_A}
\end{equation}
where $v_A$ is the Alfv\'en speed, ${\rho}_0=\mu m_p n_0$ and 
${\rho}_e = \mu m_p n_e$ are the mass density interior and exterior 
to the flux tube, $a$ is the flux tube radius, and $B_0$ and $B_e$
are the magnetic field strengths interior and exterior to the
fluxtube. 	

In Fig.3 we show the observed period ranges $P$ versus the radio
frequency ${\nu}$, as tabulated in Table II. Let us estimate what
relation between these two parameters is expected for the
fast sausage mode. We consider coronal loops within a height
range of $h=2,...,500$ Mm $(h/r_{\odot}=0,...,0.7)$. The electron
density in this height range can be estimated from the 
Baumbach-Allen model, which is in this height range (K-Corona),
\begin{equation}
	n(h) = 2.99 \times 10^8 \left( {1 \over 1+h/r_{\odot}} \right)^{16} \ q_n \quad
	[{\rm cm}^{-3}] \ ,
\end{equation}
where $q_n=10,...,100$ is a typical over-density ratio for active
region loops, compared with the background corona. We assume that
radio pulsations are emitted at a frequency 
near the local plasmafrequency ${\nu}_p$,
which is a strict function of the local electron density $n_e$,
\begin{equation}
	\nu \approx {\nu}_p = 8980 \sqrt{n_e[cm^{-3}]} \quad [{\rm Hz}] \ ,
\end{equation}
in cgs-units, and yields frequencies of $\nu \approx 20-1000$ MHz
for our coronal loop densities. For the fast sausage mode, the
loop cross-section $a$ is needed, which we assume to scale 
proportionally to the loop height $h$, say
\begin{equation}
	a = 0.2 \times h \ .
\end{equation}
Using then a canonical value for the coronal Alfv\'en speed of
$v_A \approx 1000-2000$ km s$^{-1}$, we find periods of
\begin{equation}
	P = {2 \pi a \over v_A} \approx 0.1 - 5.0 \ s \ ,
\end{equation}
In Fig.3 we show the functions $P(\nu )$ expected for
two values of the Alfv\'en velocity 
($v_A$=1000 km s$^{-1}$, and 2000 km s$^{-1}$)
and for two over-density factors $q_n=10$, and 100, which roughly
bracket the observed values $P(\nu )$. We see a tendency that
the periods become longer for low (metric) frequencies, which
probably belong to large-scale loops in the upper corona. 
All in all, the fast sausage MHD mode seems to explain the observed
periods in the range of $0.5<P<5$ s satisfactorily in the observed
frequency range of ${\nu}\approx 25-1000$ MHz. 

The fast sausage MHD mode is a radial oscillation of the flux tube
radius $a(t)$, and thus the cross-sectional area varies as
$A(t)=\pi a^2(t)$, and the loop density varies reciprocally, i.e.
$n_e(t)=n_0*A_0/A(t)\propto a^{-2}(t)$. Since the magnetic flux
$\Phi(t)=B(t)A(t)=B_0 A_0$ is conserved, the magnetic field varies
as $B(T) \propto A^{-1}(t) \propto a^{-2}(t)$. This way, the 
density and magnetic field variations will modulate any radio 
emission that is related to the plasma frequency or gyrofrequency.
Rosenberg (1970) interpreted the radio emission in terms of
gyro-synchrotron emission, in which case the emissivity scales as
\begin{equation}
	I(t) \propto n_e(t) B^2(t) \propto a^{-6}(t) \ .
\end{equation}
Therefore, the gyro-emissivity amplifies diameter variations by the
sixth power, which is perhaps the reason why radio detection of
fast sausage modes are commonly detected. The magnetic field
variation due to the sausage mode modulates also the loss-cone
angle, which can then modulate any type of (coherent) radio emission
that is driven by a loss-cone instability. Imaging observations in radio
have still insufficent resolution to determine the exact source
location, but the observations are consistent with gyro-synchrotron
sources near flare loop footpoints and with plasma emission from
type III-emitting electron beams above flare sites. Thus, fast 
sausage MHD mode oscillations can modulate the emissivity of 
nonthermal electrons in manifold ways, but progress in modeling 
can only be obtained with high-resolution imaging in many radio 
frequencies, such as with the planned future FASR instrument
(Bastian et al. 1998).

\subsection{Long-Period Radio Oscillations}

There are also reports of radio oscillations with significantly longer
periods (Table III), up to periods of an hour. Such longer periods are more
likely to be produced by standing modes that result from wave reflections
in longitudinal direction (forth and back) a coronal loop, either with
Alfv\'enic speed (fast kink mode), or with slow magneto-acoustic speed
(slow modes). The period of the fast kink mode, which corresponds to
a harmonic transverse displacement of a loop, is (Roberts et al. 1984)
\begin{equation}
	P_{fast-kink} = {2 L \over j c_k}
	= {4 \pi^{1/2} L \over j} 
	  \left( { {\rho}_0 + {\rho}_e \over
		B_0^2 + B_e^2 } \right)^{1/2}
	\approx {2 L \over j v_A}
\end{equation}
where $c_k$ is the phase speed, $L$ is the full loop length, 
and $j$ the node number ($j=1$ for fundamental harmonic number). 
Thus, for typical loop lengths of
$L=50-500$ Mm and an Alfv\'en speed of $v_A=1000$ km s$^{-1}$ we
expect kink mode periods of $P=100-1000$ s, or $P=1.4-14$ minutes. 
The fast kink mode represents just a lateral displacement of
the loop position, so it does not change the magnetic field
or electron density in first order, and thus it can not easily
be understood how it modulates radio emission. There is indeed
a paucity of detected radio periods in the range of $P=100-1000$ s
(Table III). There are essentially only three observations that 
report periods in this range, which seem to be clustered around
3 and 5 minute oscillations (Chernov et al. 1998; Gelfreikh et al.
1999; Nindos et al. 2002), which seem to be related to sunspots
and thus might be explained in terms of modulation by penumbral 
waves or global p-mode oscillations. 

\begin{table}
\begin{tabular}{llll} \hline
Observer 		& Frequency $\nu$ [MHz] & Period P[s] & Spatial scale \\
\hline
Janssens and White (1969) & 2700-15400	& 17, 23	&\\
Parks and Winckler (1969) & 15400	& 16		&\\
Janssens et al. (1973)	& 3000		& 10-20		&\\
Kaufmann (1972)		& 7000		& 2400		&\\
Kobrin and Korshunov (1972) & 9670, 9870 & 1800-3600	&\\
Tottet et al. (1979)	& 169 & 60		&6' (Nan\c{c}ay)\\
Aurass and Mann (1987)	& 23-40		& 44-234	&\\
Aschwanden et al. (1992) & 1500 	& 8.8	&0.2'-0.9' (VLA, OVRO)\\
Zlobec et al. (1992)	& 333 MHz	& 9.8-14.2	&0.7'-1.5' (VLA) \\
Baranov and Tsvetkov (1994) &8500-15,000& 22, 30, 34	&3.6'-6.0' (Crimea)\\
Qin et al. (1996)	& 2840, 9375, 15,000 & 40	&\\
Wang and Xie (1997)	& 1420, 2000	& 44, 47	&\\
Chernov et al. (1998)	& 164-407	& 180		&5'-7' (Nan\c{c}ay)\\
Gelfreikh et al. (1999) & 17,000	& 120-220	&0.2' (Nobeyama)\\
Nindos et al. (2002)	& 5000, 8000	& 157, 202 	&1.1"-5.7" (VLA)\\
Klassen et al. (2001)	& 150, 260	& 3-15		&\\
\hline
\end{tabular}
\caption{Coronal oscillations observed in radio (long periods: $P>5$ s)}
\end{table}

The slow mode, which corresponds to a bounce time of a disturbance
with the sound speed, has a period of (Roberts et al. 1984)
\begin{equation}
	P_{slow} = {2 L \over j c_T}
	= {1.2 \times 10^{-4} L \over j T_0^{1/2} } 
	  \left( 1 + { c_0^2 \over v_A^2 } \right)^{1/2}
	\approx {2 L \over j c_0}
\end{equation}
where $c_T$ is the tube speed, which is close to the sound speed 
$c_0$. Thus, for a coronal temperature of $T_0=2$ MK and loop
lengths of $L=50-500$ Mm we expect periods of $P=420-4200$ s,
or $P=7-70$ minutes. We see that the periods of the slow mode 
covers the largest observed radio periods in Table III
(Kaufmann 1972; Kobrin \& Korshunov 1972), and thus may serve 
as potential interpretation for these cases. An acoustic wave
propagating forth and back a coronal loops modulates the density,
and thus is able to modulate plasma emission as well as 
gyro-synchrotron emission.

\section{Coronal Oscillations in Optical}

\begin{table}
\begin{tabular}{llll} \hline
Observer 		& Wavelength $\lambda$ [\AA ] & Period P[s] & Instrument \\
\hline
Koutchmy et al. (1983)	& 5303		& 43, 80, 300	&Sac Peak\\
Pasachoff and Landman (1983) & 5303	& 0.5-2 (?)	&Hydrabad (eclipse)\\
Pasachoff and Ladd (1987) & 5303	& 0.5-4	(?)	&East Java (eclipse)\\
Jain and Tripathy (1998) & H$\alpha$	& 180-300	&Udaipur\\
Pasachoff et al. (2000)	& 5303		& -		&Chile (eclipse)\\
Pasachoff et al. (2002)	& 5303		& -		&Romania (eclipse)\\
Williams et al. (2001, 2002)& 5303	& 6		&Bulgaria (eclipse)\\
Ofman et al. (1997)			& 360, 1200-3000 &SOHO/UVCS, WLC \\
Ofman et al. (2000a)			& 400, 625 	&SOHO/UVCS, WLC \\
\hline
\end{tabular}
\caption{Coronal oscillations observed in optical and H$\alpha$}
\end{table}

The detection of coronal oscillations in optical wavelengths (Table IV)
seems to be
rather difficult due to the sky fluctuations in the Earth atmosphere,
but several searches were conducted, motivated by the
theoretical possibility of coronal heating by waves. Koutchmy, Zugzda,
and Locans (1983) analyzed Fe XIV green line (5303 \AA ) spectra recorded above
the solar limb and did not found any significant period in intensity,
but discovered quite significant power in the Doppler shift signal
at periods of 43, 80, and 300 s,
the latter being conincident with global p-mode oscillations. Koutchmy
et al. (1983) suggested that the Doppler velocity oscillations could be
due to resonant Alfv\'en waves. Then there were searches for high-frequency
oscillations ($P=0.5-10$ s) performed in a number of total solar eclipse
observations (Pasachoff \& Landman 1983; Pasachoff \& Ladd 1987;
Pasachoff et al. 2000, 2002), but none or only a marginal excess 
(at the $~\lapprox 1\%$ level) was found in power spectra of the
5303 \AA\ green line. Analysis of similar eclipse data in the same
wavelength with a different instrument (SECIS), however, provided
evidence for oscillations at a period of $P=6$ s (Williams et al. 2001, 
2002). Williams et al. (2001) identified the location of the coronal
oscillation in a small-scale loop (with a length of $L\approx 50$ Mm), 
and measured 
a propagation speed of $v_A \approx 2000$ km s$^{-1}$ at the location
of the oscillating signal. Based on these observables they interpret
the phenomenon as an {\sl impulsively generated fast mode wave} that
is generated at one footpoint of the loop and propagates along it.

For completeness we mention also a detection of intensity oscillations
in H$\alpha$ wavelengths (Jain and Tripathy 1998). They found prominent
5- and 3-minute modes in flares, which seem to be linked to global
p-mode oscillations. However, they find frequency deviations of about
300 $\mu$Hz between the coronal and chromospheric oscillation periods,
which they attribute to differences in the magnetic field and flare
plasma temperature. 

Quasi-periodic brightness fluctuations were also measured in polarized
brightness time series in coronal holes at heights of $\gapprox 2 R_{\odot}$
with the white-light channel of UVCS onboard SOHO, exhibiting significant
peaks in the Fourier power spectrum at $P\approx 6-10$ min, and possibly
at $P\approx 20-50$ min (Ofman et al. 1997, 2000a). These oscillation have
been interpreted in terms of slow magnetosonic waves propagating in 
coronal plumes (Ofman et al. 2000b).

\section{Coronal Oscillations in EUV}

The great breakthrough in detecting coronal oscillations came with the
availability of sub-arcsecond images in EUV, as provided by the
{\sl Transition Region And Coronal Explorer (TRACE)} since 1998. 
A compilation of EUV detections of coronal oscillations and
modeling studies is given in Table V.

\begin{table}
\begin{tabular}{llll} \hline
Observer 		& Wavelength $\lambda$ [\AA ] & Period P[s]  & Instrument \\
\hline
Chapman et al. (1972)		& 304, 315, 368	& 300 		& OSO-7		\\
Antonucci et al. (1984)	        & 554, 625, 1335& 141, 117 	& Skylab	\\
DeForest and Gurman (1998) 	& 171		& 600-900 (prop.) & SoHO/EIT	\\
Aschwanden et al. (1999)	& 171, 195	& 276$\pm$25 	& TRACE		\\
Nakariakov et al. (1999)	& 171		& 256		& TRACE		\\
Berghmans and Clette (1999)	& 195		& (prop.)	& SoHO/EIT	\\
De Moortel et al. (2000)	& 171		& 180-420 (prop.) & TRACE 	\\
Nakariakov and Ofman (2001)	& 171, 195	& 256, 360 	& TRACE		\\
Robbrecht et al. (2001)		& 171, 195	& (prop. waves) & TRACE, EIT 	\\
Schrijver et al. (2002)		& 171, 195	& -		& TRACE		\\
Aschwanden et al. (2002a)	& 171, 195	& 120-1980 	& TRACE		\\
De Mooertel et al. (2002a,b)	& 171, 195	& 282$\pm$93 (prop.)	& TRACE		\\
\hline
\end{tabular}
\caption{Coronal oscillations observed in EUV }
\end{table}

The ultimate proof of coronal loop oscillations was produced by direct imaging
of their spatial displacements, while all previous reports merely were inferred
from the periodicity in time profiles (without imaging observations). After the
1998 July 14, 12:55 UT, flare, a number of at least five loops in the flaring 
active region were discovered to exhibit periodic transverse displacements,
with an average period of $P=280\pm30$ s (Aschwanden et al. 1999). The transverse
displacement (with an amplitude of $A=4100\pm1300$ km) amounted only to a few
percent of the loop lengths ($L=130,000\pm30,000$ km). The standing wave 
could be identified as a {\sl fast kink MHD mode} in the fundamental eigen-mode, 
based on the characteristics of fixed nodes (near the footpoints of the loop) 
and the lateral displacements (near the apex of the loop). Moreover, the
observed period $P$ matched the theoretical expression for the fast kink mode
$P_{fast-kink}\approx 2 L / v_A$ (Eq.7), based on the observed loop lengths $L$ 
and estimated Alfv\'en speeds $v_A$. In Fig.4 we show the probabilities for
different MHD modes based on physical parameters ($L, n_e, B$) that were
measured from similar loops (Aschwanden et al. 1999). However, although
the notion of {\sl fast kink MHD mode} oscillations may be correct in first
order, observations often show substantial deviations from a nice symmetric
kink mode oscillation, asymmetric excitation, and systematic eigen-motion
trends superimposed on the oscillatory displacements (Aschwanden et al. 2002a). 

\begin{figure}[t]
\centerline{\epsfig{file=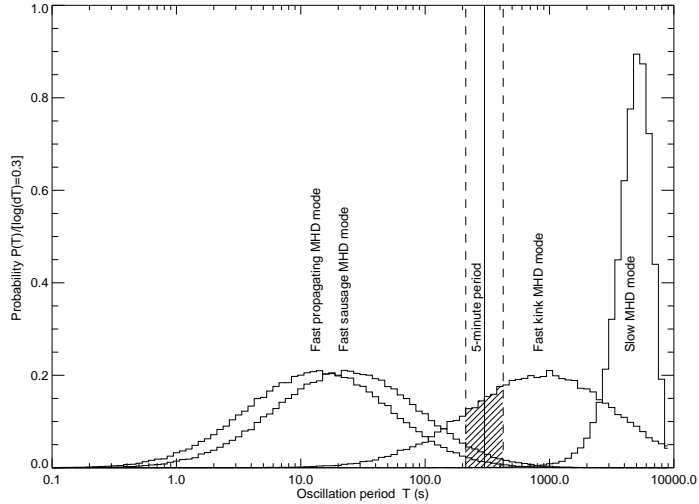,width=\textwidth}} 	
\caption{Probability distributions of four different MHD periods calculated for
densities, magnetic fields, and altitudes that are typical for EUV
loops observed in the temperature range of $T_e=1.0-1.5$ MK (e.g. with
TRACE or SoHO/EIT 171 \ang ). The probabilities are normalized for
period bandwidths covering a factor of 2, i.e. per log(dT)=0.3. Note
that the fast kink MHD mode has the highest probability of coinciding
with the 5-minute period [Aschwanden et al. 1999].}
\end{figure}

Another interesting property is the strong damping, which brings loops
oscillations to a halt after a few periods. For instance, the damping time 
of an oscillating loop with a period of $P=261$ s (Fig.5) was measured to
$t_D=1183$ s (Aschwanden et al. 2002a), and independently as $P=256\pm8$ s
and $t_d=870$ s (Nakariakov et al. 1999), so it corresponds only to about
3 oscillation periods. A larger statistics of some 26 loops confirmed the
general trend of strong damping, with an average of $t_D/P=4.0\pm1.8$ periods.
There are at least five different theories considered for the loop damping:
(1) nonideal effects (viscous and ohmic damping, optically thin radiation,
thermal conduction), (2) wave leakage through the sides of the loops,
(3) wave leakage at the footpoints, (4) phase mixing, and (5) Alfv\'enic
resonant absorption. The first two damping effects are considered to be
too inefficient to explain the observed loop damping (Roberts 2000).   
Wave leakage at footpoints (DePontieu et al. 2001) is a too weak damping
force for standard chromospheric scale heights (Ofman 2002), 
say ${\lambda}_H \approx 500$ km, but could account for slightly extended
chromospheric scale heights, say a factor of $\approx 2$ (Aschwanden et al.
2002a). Evidence for the existence of an {\sl extended (spicular) chromosphere} 
is supported by sub-mm radio observations of the solar limb (Ewell et al. 1993)
as well as by recent {\sl RHESSI} observations (Aschwanden, Brown, and Kontar
2002b). Phase mixing (Heyvaerts \& Priest 1983), which provides a viscous damping
mechanism due to the differences in Alfv\'en speeds in an inhomogeneous medium,
was found to yield a scaling law of the damping time with loop period that
is consistent with observations, if 
the length scale of inhomogeneity is assumed to be proportional to the loop
length (Ofman \& Aschwanden 2002). On the other side, Alfv\'enic resonant
absorption (Ionson 1978; Rae \& Roberts 1982)  
provides also an efficient damping mechanism if the cross-sectional density
gradient at the loop edges is sufficiently flat (Ruderman \& Roberts 2002; 
Goossens, Andries, \& Aschwanden 2002). Nakariakov et al. (1999) concluded
that the coronal value of the viscous and resistive dissipation 
coefficient could be 8-9 orders of magnetiude larger than the classical value.
Such high Reynolds numbers are actually in line with fast reconnection rates
estimated in the corona (Dere 1996). So the jury is still out which physical 
mechanism is dominant in the strong loop damping.

\begin{figure}[t]
\centerline{\epsfig{file=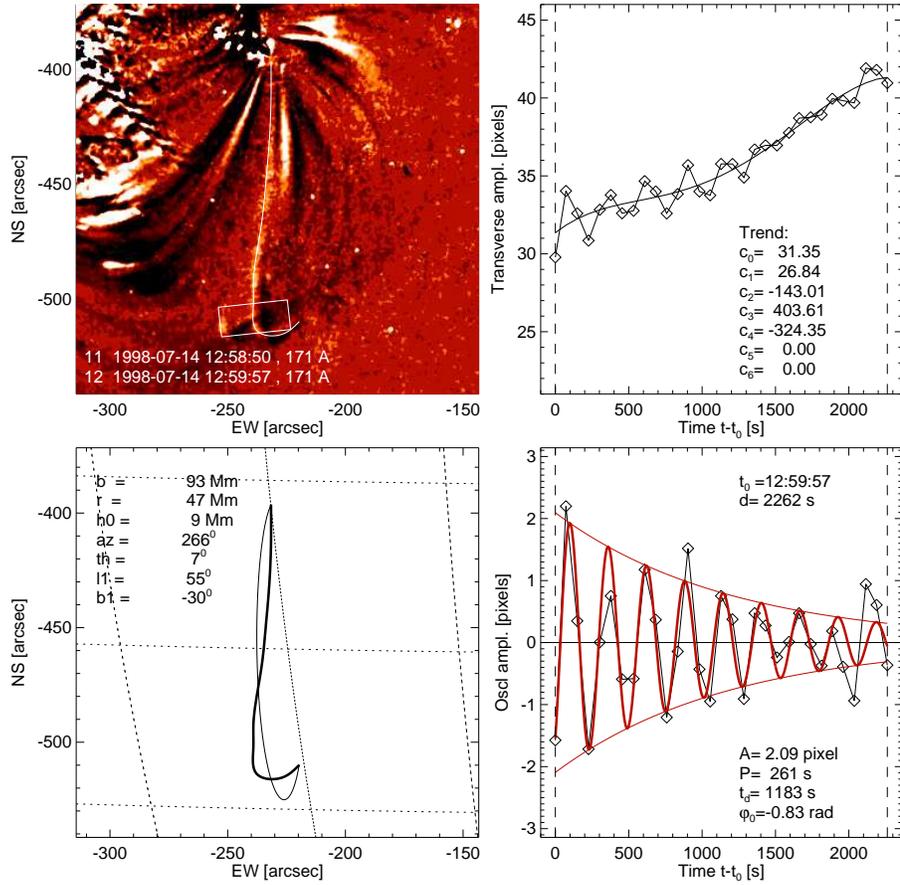,width=\textwidth}} 	
\caption{Oscillation event on 1998-Jul-14, 12:45 UT.
This loop is identical with the case analyzed in Nakariakov et al. (1999).
{\sl Top left:} Difference image with rectangular box indicates where transverse oscillations are
analyzed.  {\sl Bottom left:} 3D geometry of observed loop (thick line) fitted with a
circular model (thin line), specified by the
baseline (b), loop radius (r), height of circular loop center
($h_0$), azimuth angle of baseline (az $=\alpha$), inclination angle of loop plane to vertical (th$=\theta$),
heliographic longitude (l1) and latitude (b1) of baseline midpoint. The spacing of the heliographic
grid is $5^0$ ($\approx 60$ Mm).
{\sl Top right:} Transverse loop position (in pixel units), where the trend is fitted
with a polynomial (with coefficients $c_0, c_1, c_2, c_3, ...$).
{\sl Bottom right:} Detrended  oscillation, fitted with an exponentially decaying
oscillatory function (with amplitude $A$, period $P$, decay time $t_d$, phase ${\varphi}_0$).
$t_0$ is the start time (to which the phase ${\varphi}_0$ is referenced),
and $d$ is the duration of the fitted interval [Aschwanden et al. 2002a].}
\end{figure}

\begin{table} %
\begin{tabular}{lrr}
\hline
Parameter           & Average & Range     \\
\hline
Loop half length $L$       & $110\pm53$ Mm      & 37-291 Mm          \\
Loop width $w$             & $8.7\pm2.8$ Mm     & 5.5-16.8 Mm        \\
Oscillation period $P$     & $321\pm140$ s      & 137-694 s $^2$)    \\
                           & $5.4\pm2.3$ min    & 2.3-10.8 min$^2$)  \\
Decay time $t_d$           & $580\pm385$ s      & 191-1246$^3$) s    \\
                           & $9.7\pm6.4$ min    &  3.2-20.8 min$^3$) \\
Oscillation duration $d$   & $1392\pm1080$ s    &  400-5388 s        \\
                           & $23\pm18$ min      &  6.7-90 min        \\
Oscillation amplitude $A$  & $2200\pm2800$ km   & 100-8800 km        \\
Number of periods          & $4.0\pm1.8$        & 1.3-8.7            \\
Electron density of loop $n_{loop}$ & $(6.0\pm3.3) 10^8$ cm$^{-3}$ &
                                      $(1.3-17.1) 10^8$ cm$^{-3}$    \\
Maximum transverse speed $v_{max}$ &  $42\pm53$ km/s & 3.6-229 km/s  \\
Loop Alfv\'en speed $v_A$    & $2900\pm800$ km/s & 1600-5600 km/s      \\
Mach factor $v_{max}/v_{sound}$& $0.28\pm0.35$   & 0.02-1.53         \\
Alfv\'en transit time $t_A$  & $150\pm64$ s        & 60-311 s          \\
Duration/Alfv\'enic transit $d/t_A$      & $9.8\pm5.7$ & 1.5-26.0      \\
Decay/Alfv\'enic transit $t_d/t_A$ & $4.1\pm2.3$ & 1.7-9.6$^3$)  \\
Period/Alfv\'enic transit   $P/t_A$ & $2.4\pm1.2$      & 0.9-5.4$^2$)  \\
\hline
\end{tabular}
\caption{Average and ranges of physical parameters of 26 oscillating EUV loops$^1$}
$^1$) All Alfv\'enic speeds and times are calculated for a magnetic field of
$B=30$ G. \\
$^2$) An extreme period of $P=2004$ s is excluded in the statistics. \\
$^3$) Only the 10 most reliable decay times $t_d$ are included in the statistics. 
\end{table}

The exciter mechanism of loop oscillations is another interesting topic.
In virtually all cases there seems to be a flare or a filament destabilization
involved (Schrijver et al. 2002). Moreover, the local magnetic field topology
at the footpoints of oscillating loops seems to have a preference for magnetic
separatrices, which would explain the amplification of oscillatory excitements
(Schrijver et al. 2002) and was simulated with magnetic field models
(Schrijver \& Brown 2000). The exciter mechanism has severe consequences for
the geometry and dynamics of the loop oscillations. An asymmetric exciter
hits a loop first at some location away from the midpoint, and thus can excite
higher harmonic nodes or a combination of multiple modes (sausage, kink,
torsional, slow mode). Because strong damping occurs, the loops do not have
enough time to settle into an eigen-frequency, and thus phase shifts of
propagating waves and complicated patterns of multiple modes occur that are
difficult to analyze.   

A summary of observed physical parameters is given in Table VI, quantifying
also the ratios of periods, damping times, and durations of observed oscillations.
If we use a canonical value of $B\approx 30$ G to estimate the Alfv\'en speed,
we find that the observed periods exceed the expected kink mode period by a
factor of 2.4 in the average. It is not clear whether the magnetic field in
the oscillating loops is lower (i.e., $B = 13\pm9$ G; Nakariakov \& Ofman 2001)
than expected from standard coronal models,
or whether the {\sl fast-kink mode period} needs to be corrected for propagating  
(impulsively generated) wave modes or for a combination of multiple modes.
Nevertheless, the high-quality images of {\sl TRACE} provide a substantial
number of physical parameters for the oscillating loops that translate into quite
rigorous constraints for theoretical models of MHD oscillations and waves. 

\begin{table}[t]
\begin{tabular}{llll} \hline
Observer 		& Wavelength $\lambda$ & Period P[s]  & Instrument \\
\hline
Jakimiec and Jakimiec (1974) & 1-8 \AA	& 200-900		& SOLRAD 9	\\
Harrison (1987) 	& 3.5-5.5 keV	& $>$1440  		& SMM/HXIS	\\
Thomas et al. (1987)	& 2-8, 8-16 \AA	& 1.6 			& OSO-7		\\
Svestka (1994)		& 0.5-4, 1-8 \AA & 1200			& GOES		\\
McKenzie and Mullan (1997) & 3-45 \AA	& 9.6-61.6		& Yohkoh/SXT	\\
Wang et al. (2002a,b)	& (Fe XIX) 1118 \AA & 660-1860		& SoHO/SUMER	\\
\hline
\end{tabular}
\caption{Coronal oscillations observed in Soft X-rays }
\end{table}

\section{Coronal Oscillations in Soft X-Rays}

There were only very few reports on oscillating loops in soft X-rays in the past
(Table VII),
one with a very fast period (Thomas et al. 1987), and some with very long periods 
(Jakimiec \& Jakimiec 1974; Harrison 1987; Svestka 1994). 
A systematic search for oscillating loops in soft X-rays with {\sl Yohkoh/SXT} data
was conducted by McKenzie
and Mullan (1997), but only 16 out of 544 cases were found to exhibit significant
periodicities, with periods in the range of $P=9.6-61.6$ s. These oscillations
were derived from fluctuations in the soft X-ray flux time profiles, which are 
proportional to the emission measure (or squared density). Density modulations 
could be most easily explained in terms of the {\sl fast sausage MHD mode} (Fig.4), 
because the {\sl fast kink MHD mode} does not cause density modulations in first order.
Also, the required Alfv\'en speeds appear to be unusually high for the {\sl fast
kink MHD mode} interpretation, as it was suggested by McKenzie and Mullan (1997).  

A renaissance of loop oscillations in soft X-rays was initiated with recent
{\sl SUMER/SoHO} observations. Searches for waves in {\sl SUMER} data were performed 
with both line width measurements (Erd\'elyi et al. 1998) as well as with Doppler
shift measurements (Wang et al. 2002a,b). As for the case of optical
observations, oscillatory signals were detected more pronounced in Doppler shifts 
than in line intensity. A total of 17 loop oscillation events have been
detected with SUMER in Fe XIX Doppler shifts, at temperatures of $T>6$ MK, although 
the majority were not associated with flare-like activity. The Doppler oscillations have
periods of $P=14-18$ minutes and exponential decay times of ${\tau}_D=12-19$ minutes,
so the ratio is only ${\tau}_D/P\approx 1$. The long periods (compare with Fig.4)
as well as the measurements of Doppler shifts support an interpretation in terms
of slow MHD modes (Eq.8). We suspect that a flare or filament destabilization causes
a pressure disturbance at one side of a loop, which propagates as a slow mode
magnetosonic (acoustic) wave in longitudinal direction along the loop and becomes 
reflected at the opposite side (Nakariakov et al. 2000). 
The rapid damping of the propagating wave seems to be
caused by thermal conduction, as simulated with an MHD code (Ofman \& Wang 2002).

\begin{table}[t]
\begin{tabular}{llll} \hline
Observer 		& Wavelength $\lambda$ [\AA ] & Period P[s]  & Instrument \\
\hline
Parks and Winckler (1969) & 		& 16 		& Balloon	\\
Lipa (1978)		&		& 10-100	& OSO-5		\\
Takakura et al. (1983)	&		& 0.3		& Hinotori	\\
Kiplinger et al. (1982)	&		& 0.4, 0.8	& SMM/HXRBS	\\
Kiplinger et al. (1983)	&		& 8.2		& SMM/HXRBS	\\
Desai et al. (1987)	&		& 2-7		& Venera	\\
Terekhov et al. (2002)	&		& 143.2$\pm$0.8	& GRANAT	\\
Asai et al. (2002)	&		& 6.6 s		& Yohkoh/HXT	\\
\hline
\end{tabular}
\caption{Coronal oscillations observed in Hard X-rays }
\end{table}

\section{Coronal Oscillations in Hard X-Rays}

Reports of oscillation detections in hard X-ray wavelengths are listed in Table VIII.
Because it has been shown that radio pulsations are often correlated with the
hard X-ray flux (Correia and Kaufmann 1987; Aschwanden, Benz, and Kane 1990; 
Kurths et al. 1991), it is likely that the
hard X-ray flux is modulated by the same quasi-periodic acceleration mechanism
that produces the radio-emitting nonthermal electrons (Aschwanden et al. 1994; 1995b). 
Therefore, the same interpretation may hold for fast hard X-ray pulsations
(e.g. Takakura et al. 1983) as for fast radio pulses (Sections 2.1 and 2.2).  
In general, however, much less fast (sub-second) pulses were detected in hard
X-rays (e.g. Kiplinger et al. 1982) than in radio, because of the poorer sensitivity 
of earlier hard X-ray instruments. For the 7 recurrent hard X-ray pulses 
(with a mean period of 8.2 s) observed during the 1980 June 7 flare (Kiplinger et al. 1983),
a detailed dynamic model of an oscillating current sheet was developed
by Sakai \& Ohsawa (1987) and Tajima et al. (1987). In this scenario, 
a current sheet undergoes oscillatory dynamics driven by pressure
balance oscillations between the lateral plasma inflow and the internal
currents that cause the current sheet collapse. This model even reproduced
the observed double-peak substructure of the quasi-periodic pulses.
On the other side, large flares always show long arcades of postflare loops,
which suggests that a spatial fragmentation of the energy release (or
intermittent bursty reconnection along the neutral line) causes a temporal
sequence of quasi-periodic or random-like hard X-ray pulses, which is difficult
to reconcile with the oscillatory dynamics of a single compact current sheet.  
So we have to await better spatial information in hard X-rays, such as with
{\sl RHESSI}, to sort out whether quasi-periodic hard X-ray pulses come from
spatially separated structures or from the same compact region.

The {\sl fast sausage MHD mode} causes a cross-sectional vibration of the
flare and modulates in this way the density and magnetic field. Brown \& McClymont 
(1976) (and similarly Zaitsev and Stepanov 1982)
suggested a betatron acceleration model in a vibrating flare loop and explained
in this way an observed correlation between the electron flux and spectral index
in one event. Such a model can explain quasi-periodic hard X-ray emission with
periods amenable to the {\sl fast sausage MHD mode} (e.g. Parks and Winckler 1969;
Lipa 1978; Kiplinger et al. 1983; Desai et al. 1987). 

For the longest periods reported in hard X-rays (e.g. P=143 s, Terekhov et al. 2002),
we might envision an interpretation in terms of {\sl slow MHD modes}, such as for
the {\sl SUMER/SoHO} observed cases of Doppler shift oscillations. 

\section{Conclusions}

The detection of oscillations and waves has grown exponentially over the
last few years, especially with space-borne EUV telescopes, which were equipped with
improved sensitivity, higher spatial resolution, and with faster cadences. Not
only the volume of observations increased, but also progress was advanced 
in the physical interpretation. 
The observations show evidence for a variety of MHD oscillation modes,
which roughly can be discriminated by their characteristic period
range for coronal conditions (see Fig.4): 
{\sl fast sausage MHD mode} ($P\approx 0.5-5$ s),
{\sl fast kink MHD mode} ($P\approx 3-7$ minutes), and 
{\sl slow (acoustic) modes} ($P\approx 10-30$ minutes).
The fastest MHD mode, i.e., the {\sl sausage mode}, could only be
detected in radio so far, and in optical perhaps (Williams et al. 2002),
where sufficiently high (sub-second) 
time resolution is available, but it could not have been 
detected with TRACE or SUMER because of the insufficient
time cadence (in the order of minutes).
The fast kink mode has definitely been detected with TRACE in EUV,
based on the lateral displacements and expected periods. The full
picture, however, may be more complicated because of asymmetric
excitation and propagating waves.
The slow (acoustic) mode seems to be detected with SUMER, based
on the periods ($P\approx 10-30$ minutes) and Doppler velocities.

The observations of oscillating systems in the solar corona raise
interesting questions about their exciting and quenching mechanisms.
The excitation of loop oscillations seems to be accomplished by
nearby flares or destabilizing filaments, with a preference near
magnetic separatrices.
Asymmetric excitation may excite multiple modes, higher harmonics,
and superimposed eigen-motion of the loop centroids.
The damping of loop oscillations is strong ($t_D/P \approx 1-3$).
Viable physical mechanisms are chromospheric leakage (for an extended
chromosphere), phase mixing (for inhomogeneous loops), and Alfv\'enic 
resonance absorption (if the loop cross-sectional density profile 
has a ``smooth'' edge). 

Besides the regular oscillations that is an intrinsic characteristic
of MHD modes, there are also aperiodic or quasi-periodic phenomena
which call for a different interpretation.
Quasi-periodic oscillations observed in radio and hard X-rays do not
have the regular periodicity as MHD oscillations and are likely to
be produced by oscillatory regimes (limit cycles) of nonlinear 
dissipative systems, e.g. in the magnetic reconnection and particle
acceleration region.

\begin{acknowledgements}
{\footnotesize  		
Part of this work has been supported by NASA contracts NAS5-38099 (TRACE) 
and NAS8-00119 (Yohkoh/SXT). Support by the NATO Science Programm for
participation at the Advanced Research Workshop is acknowledged. 
The author thanks Bernie Roberts, Robertus Erd\'elyi, Valery Nakariakov,
Leon Ofman, Tongjang Wang, Werner Curdt, and the referee for instructive 
and helpful discussions and comments. }
\end{acknowledgements}

\end{article}
\end{document}